\documentclass[twoside,leqno,twocolumn]{article}

\usepackage[letterpaper]{geometry}

\usepackage{ltexpprt}

\newtheorem{mydef}{Definition}
\newtheorem{myrem}{Remark}
\newtheorem{myeg}{Example}
\newcommand{\bKL}{\mathbf{KL}}

\newcommand{\bPr}{\mathbf{Pr}}

\usepackage{amssymb,amsmath}
\usepackage[algo2e,linesnumbered,ruled,vlined,commentsnumbered]{algorithm2e}
\usepackage{enumitem}
\usepackage[disable]{todonotes}
\usepackage{caption}
\usepackage{subcaption}
\usepackage{multirow}
\usepackage[normalem]{ulem}
\usepackage{array, booktabs}

\usepackage{hyperref}
\newcommand{\ptitle}[1]{{\bf #1.}}
\newcolumntype{M}[1]{>{\centering\arraybackslash}m{#1}}
\newcolumntype{P}[1]{>{\arraybackslash}p{#1}}

\setlist[itemize]{noitemsep, topsep=0pt, leftmargin=11pt}

\begin{document}

\title{\Large Explainable Subgraphs with Surprising Densities: A Subgroup Discovery Approach\thanks{This research was funded by the ERC under the EU's Seventh Framework Programme (FP7/2007-2013) / ERC Grant Agreement no. 615517, the Flemish government under the ``Onderzoeksprogramma Artificiële Intelligentie (AI) Vlaanderen" programme, the FWO (project no. G091017N, G0F9816N), and from the EU's Horizon 2020 research and innovation programme and the FWO under the Marie Sklodowska-Curie Grant Agreement no. 665501.}}
\author{Junning Deng \thanks{IDLab, Ghent University; Firstname.Lastname@UGent.be}
\and Bo Kang \footnotemark[2]
\and Jefrey Lijffijt \footnotemark[2]
\and Tijl De Bie  \footnotemark[2]}

\date{}
%
\maketitle







\begin{abstract} \small\baselineskip=9pt 
The connectivity structure of graphs is typically related to the attributes of the nodes. In social networks for example, the probability of a friendship between two people depends on their attributes, such as their age, address, and hobbies.
The connectivity of a graph can thus possibly be understood in terms of patterns of the form `the subgroup of individuals with properties X are often (or rarely) friends with individuals in another subgroup with properties Y'
. Such rules present potentially actionable and generalizable insights into the graph.
We present a method that finds pairs of node subgroups between which the edge density is interestingly high or low, using an informa\-tion-theoretic definition of interestingness. This interestingness is quantified subjectively, to contrast with prior information an analyst may have about the graph. This view immediately enables iterative mining of such patterns.
Our work generalizes prior work on dense subgraph mining (i.e. subgraphs induced by a \emph{single} subgroup). Moreover, not only is the proposed method more general, we also demonstrate considerable practical advantages 
for the single subgroup special case.
\end{abstract}

\section*{Keywords} 
Graph mining, Subgroup Discovery, Subjective interestingness, Community detection

\section{Introduction\label{sec:introduction}}

Real-life graphs (\emph{aka} networks) often contain attributes for the nodes.
In social networks for example, nodes correspond to individuals and node attributes can include their age, address, hobbies, etc.
A network's connectivity is usually related to those attributes:
individuals' attributes affect the likelihood of them meeting, and, if they meet, of becoming friends.
Hence, to a certain extent, it should be possible to understand the connectivity of a graph in terms of those attributes.

One approach to identify the relations between the connectivity and the attributes is to train a link prediction classifier, with as input the attribute values for a pair of nodes, and predicting the edge as present or absent.
Such global models often fail to provide insight though, much like a global classifier on any data type may fail to provide insight in other classification problems.
To address this, the local pattern mining community introduced the concept of \emph{subgroup discovery}, which aims to identify subgroups of data points for which a target attribute has homogeneous and/or outstanding values.
Subgroups are local patterns, in that they provide information only about a certain part of the data.

Research on local pattern mining in attributed graphs has so far focused on identifying dense node-induced subgraphs, dubbed \emph{communities}, that are coherent also in terms of attributes.
There are two complementary approaches.
The first explores the space of communities that meet certain criteria in terms of density, in search for those that are homogeneous.
The second explores the space of rules over the attributes, in search for those that define subgroups of nodes that form a dense community. This is effectively a subgroup discovery approach to dense subgraph mining.
%

\ptitle{Limitations of the state-of-the-art}
Both of these approaches make use of attribute homophily: the tendency of links to exist between nodes sharing similar attributes.
While the homophily assumption is often reasonable, it also limits the scope of application of prior work to finding dense communities with homogeneous attributes.
A \emph{first limitation} of the state-of-the-art is thus its inability to find e.g. sparse subgraphs.

A \emph{second limitation} is that the interestingness of such patterns has invariably been quantified by objective measures---i.e. measures independent of the data analyst's prior knowledge. Yet, the most `interesting' patterns found are often obvious and implied by such prior knowledge (e.g. communities involving high-degree nodes, or in a student friendship network, communities involving individuals practicing the same sport), making them subjectively uninteresting.
%

A \emph{third limitation} of prior work is that the patterns describe only the connectivity \emph{within} communities and not \emph{between} subgroups of nodes. As an obvious example, this excludes patterns that describe friendships between a particular subgroup of female and a subgroup of male individuals in a social network. The experiments on real-life networks contain many less obvious examples.


\ptitle{Contributions}
We depart from the existing literature in formalizing a \emph{subjective} interestingness measure, building on the ideas from the FORSIED framework \cite{debie2011a
}, and this for \emph{sparse} as well as for \emph{dense} subgraph patterns. In this way, we overcome the first and second limitations of prior work discussed above. 
Moreover, this interestingness measure is naturally applicable for patterns describing the graph density between a pair of subgroups, to which we will refer as \emph{bi-subgroup patterns}. Hence, our method overcomes the third limitation of prior work.
Our specific contributions are:
(1) Novel definitions of single-subgroup patterns and bi-subgroup patterns [Sec.~\ref{sec: pattern syntax}].
(2) A formalization of \emph{Subjective Interestingness} (SI), based on the analyst's evolving prior beliefs [Sec.~\ref{sec: SI}].
(3) A beam-search algorithm to mine the subjectively most interesting bi-subgroup patterns [Sec.~\ref{sec: algorithm}].
(4) An empirical evaluation on real-world data, confirming our method's ability to 
identify subjectively interesting patterns [Sec.~\ref{sec: experiments}].

\section{Subgroup pattern syntaxes for graphs\label{sec: pattern syntax}}
This section formalizes single-subgroup and bi-subgroup patterns for graphs, beginning with some notation.

An attributed graph is denoted as a triplet $G=(V, E, A)$ where $V$ is a set of $n=|V|$ vertices, and $E\subseteq\binom{V}{2}$ is a set of $m=|E|$ undirected edges\footnote{We consider undirected graphs for the sake of presentation and consistency with most literature. However, we note that all our results can be easily extended to directed graphs and graphs with self-edges.}, and $A$ is a set of attributes $a\in A$ defined as functions $a:V\rightarrow \mbox{Dom}_a$, where $\mbox{Dom}_a$ is the set of values the attribute can take over $V$.
%
%
For each attribute $a\in A$ with nominal $\mbox{Dom}_a$ and for each $y\in\mbox{Dom}_a$, we introduce a Boolean function $s_{a,y}: V \to\{\text{true},\text{false}\}$, defined as true for $v\in V$ iff $a(v)=y$. Analogously, for each $a\in A$ with real-valued $\mbox{Dom}_a$ and for each $l<u$ and $l,u\in\mbox{Dom}_a$, we define $s_{a,[l,u]}: V \to\{\text{true},\text{false}\}$, with $s_{a,[l,u]}(v)\triangleq\text{true}$ iff $a(v)\in[l,u]$. We call these functions \emph{selectors}, and denote the set of all selectors as $S$. A \emph{description} or \emph{rule} $W$ is a conjunction of a subset of selectors: $W = s_1\wedge s_2\ldots \wedge s_{|W|}$. The \emph{extension} $\varepsilon(W)$ of a rule $W$ is defined as the subset of vertices that satisfy it: $\varepsilon(W) \triangleq \{v\in V | W(v) = \text{true}\}$. We informally also refer to the extension as the \emph{subgroup}. Now a \emph{description-induced subgraph} can be formally defined as:

\begin{mydef}
\emph{(Description-induced-subgraph)} Given an attributed graph $G=(V,E,A)$ and a description $W$, we say that a subgraph $G[W]=(V_W,E_W,A)$ where $V_W \subseteq V, E_W \subseteq E$, is induced by $W$ if:
  \begin{enumerate}[label=(\roman*)]
  \item
    $V_W=\varepsilon(W)$, i.e., the set of vertices from $V$ that is the extension of the description $W$, and
  \item
    $E_W=(V_W\times V_W) \cap E$, i.e., the set of edges from $E$ that have both endpoints in $V_W$.
  \end{enumerate}
\end{mydef}

\begin{figure}
 \begin{minipage}[b]{0.5\textwidth}
  	\centering
	\includegraphics[width=7cm,height=2.5cm]{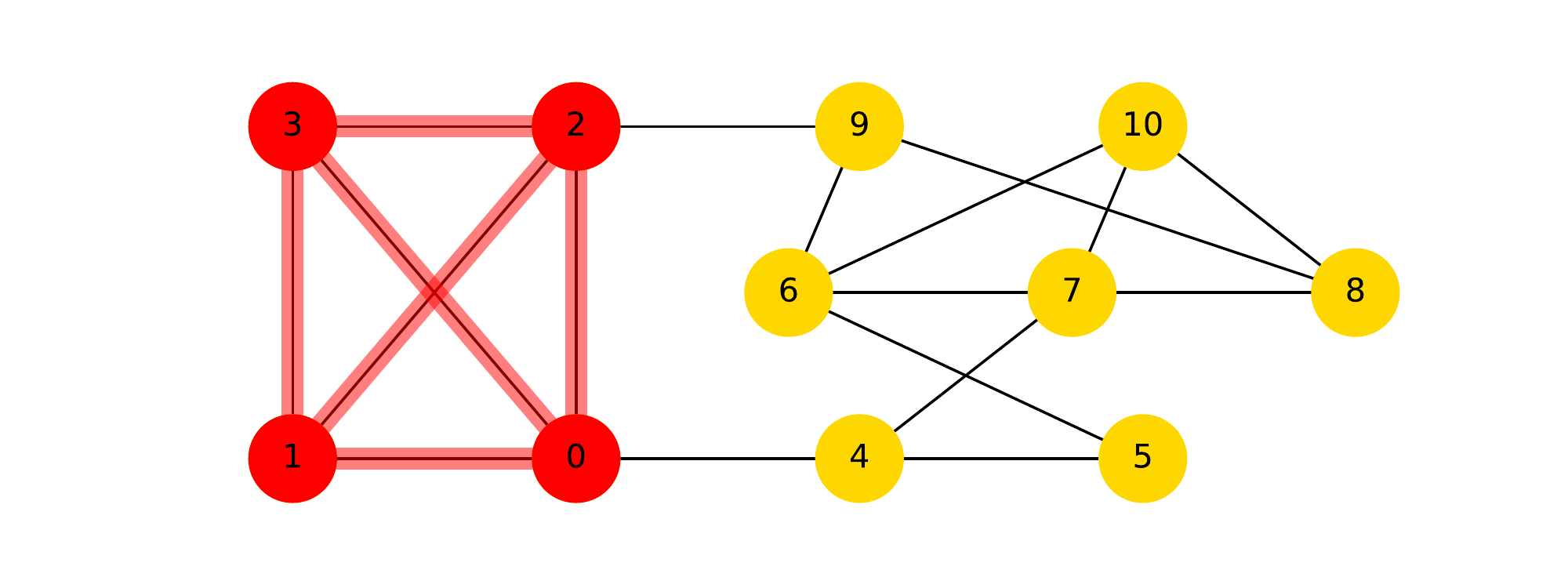}
	\caption*{(a) Graph} 
	\label{fig: eg_nw}	
 \end{minipage}
  \hfill
  \begin{minipage}[b]{0.5\textwidth}
    \centering
 \resizebox{\textwidth}{!}{
    \begin{tabular}{c|ccccccccccc}\hline
    	 \textbf{Vertex}&\textbf{0}&\textbf{1}&\textbf{2}&\textbf{3}&\textbf{4}&\textbf{5}&\textbf{6}&\textbf{7}&\textbf{8}&\textbf{9}&\textbf{10}  \\ \hline
	 $\mathbf{a}$&3.5&2.6&3.8&3.2&1.8&1.2&5.4&0.9&6.7&2.3&3.1\\\hline
	 $\mathbf{b}$&1&1&1&1&1&0&0&1&0&0&0\\\hline
	 $\mathbf{c}$&0&0&1&0&0&0&1&1&1&1&1\\\hline
	 $\mathbf{d}$&1&0&1&1&1&0&0&0&0&1&0\\ 
        \hline
      \end{tabular}}
	
      \captionof*{table}{(b) Vertex attributes}
      \label{tab: attributes}
    \end{minipage}
    \caption{Example attributed graph with $11$ vertices (0-10) and $4$ associated attributes ($a$-$d$). The subgraph induced by the description ($W=s_{a,[2,4]}\wedge s_{b,1}$) is highlighted in red.}
    \label{fig: example}
  \end{figure}

\begin{myeg}
Fig.~\ref{fig: example}(a) displays an example attributed graph with $11$ vertices, $18$ edges
. Each node is annotated with $1$ real-valued attribute ($a$) and $3$ binary attributes ($b,c,d$). Consider a description $W=s_{a,[2,4]}\wedge s_{b,1}$. The extension of this description is the set of nodes with attribute $a$ value from $2$ to $4$ and attribute $b$ as 1, i.e., $\varepsilon(W)=\{0,1,2,3\}$. The subgraph induced by $W$ is formed from $\varepsilon(W)$ and all the edges connecting pairs of vertices in that set (highlighted with red in Fig.~\ref{fig: example}(a)). 
\end{myeg}

\subsection{Single-subgroup pattern}

A first pattern syntax we consider informs the analyst about the density of a description-induced subgraph $G[W]$. We assume the analyst is satisfied by knowing whether the density is unusually small, or unusually large, and given this does not expect to know the precise density. It thus suffices for the pattern syntax to indicate whether the density is either smaller than, or larger than, a specified value.
We thus formally define the \emph{single-subgroup} pattern syntax as a triplet $(W,I,k_W)$, where $W$ is a description and $I\in\{0,1\}$ indicates whether the number of edges $E_W$ in subgraph $G[W]$ induced by $W$ is greater (or less) than $k_W$. Thus, $I=1$ indicates the induced subgraph is sparse, whereas $I=0$ characterizes a dense subgraph. The maximum number of edges in $G[W]$ is denoted by $n_W$, equal to $\frac{1}{2}|\varepsilon(W)|(|\varepsilon(W)|-1)$ for undirected graphs without self-edges.

\subsection{Bi-subgroup pattern}

We also define a pattern syntax informing the analyst about the edge density between two different subgroups. More formally, we define a \emph{bi-subgroup pattern} as a quadruplet $(W_1,W_2,I,k_W)$, where $W_1$ and $W_2$ are two descriptions, and $I\in\{0,1\}$ indicates whether the number of connections between $\varepsilon(W_1)$ and $\varepsilon(W_2)$ is upper bounded (1) or lower bounded (0) by the threshold $k_W$. The maximum number of connections between the extensions $\varepsilon(W_1)$ and $\varepsilon(W_2)$ is denoted by  $n_{W}\triangleq |\varepsilon(W_1)||\varepsilon(W_2)|-\frac{1}{2}|\varepsilon(W_1\land W_2)|(|\varepsilon(W_1\land W_2)|+1)$ for undirected graphs without self-edges. Note that single-subgroup patterns are a special case of bi-subgroup patterns when $W_1\equiv W_2$.

%

\begin{myrem}
Although $k_W$ for a pattern $(W_1,W_2,I,k_W)$ can be any value with which the number of connections between $\varepsilon(W_1)$ and $\varepsilon(W_2)$ (or within $\varepsilon(W_1)$ when $W_1\equiv W_2$) are bounded, our work focus on identifying patterns whose $k_W$ is the actual number of connections between these two subgroups (or within this single subgroup when $W_1\equiv W_2$), as such patterns are maximally informative. 
\end{myrem}

\section{Formalizing the subjective interestingness \label{sec: SI}}

Previous work on mining patterns in attributed graphs focuses on identifying dense communities, with \emph{density} quantified in an objective way (see Sec.~\ref{sec: relatedwork}). However, given prior information on the graph, the resulting patterns may be trivial, containing limited information that is novel to the analyst.
Tackling this necessitates the use of subjective measures of interestingness.

\subsection{General approach}

We follow the approach as outlined by De Bie \cite{debie2011b} to quantify the SI of a pattern
. In this framework, the analyst's belief state is modeled by a \emph{background distribution} over the data space.
This background distribution represents any prior beliefs the analyst may have by assigning a probability (density) to each possible value for the data according to how plausible the analyst thinks this value is. 
As such, the background distribution also makes it possible to 
assess the surprise in the analyst when informed about the presence of a pattern.
It was argued that a good choice for the background distribution is the maximum entropy distribution subject to some particular constraints that represent the analyst's prior beliefs about the data.
As the analyst is informed about a pattern, the knowledge about the data will increase, and the background distribution will change. For details see Sec.~\ref{sec:bd}.

Given a background distribution, the SI of a pattern can be quantified as the ratio of the \emph{Information Content} (IC) and the \emph{Description Length} (DL) of a pattern. The IC is defined as the amount of information gained when informed about the pattern's presence, computed as the negative log probability of the pattern w.r.t. the background distribution $P$. 
The DL quantifies the code length needed to communicate the pattern to the analyst. These are discussed in more detail in Sec.~\ref{sec:si}, but first we further explain the background distribution.

\subsection{The background distribution}\label{sec:bd}

\subsubsection*{The initial background distribution}
Here we recapitulate how prior beliefs of the following types can be modelled in a background distribution: (i) on individual vertex degrees; (ii) on the overall graph density; (iii) on densities between bins.

\textit{Type (i) and (ii): Prior beliefs on individual vertex degrees and on the overall graph density.}
Given prior beliefs about the degree of each vertex, the maximum entropy distribution is a product of independent Bernoulli distributions, one for each of the random variable $h_{u,v}$ defined as $1$ if $(u,v)\in E$ and $0$ otherwise \cite{debie2011b}. Denoting the probability that $h_{u,v}=1$ by $p_{u,v}$, this distribution is of the form:
\begin{align*}
P(E)&=\prod_{u,v} {p_{u,v}}^{h_{u,v}}\cdot (1-p_{u,v})^{1-h_{u,v}},\\
\text{where}\quad p_{u,v}&= \frac{\exp(\lambda^{r}_{u}+\lambda^{c}_{v})}{1+\exp(\lambda^{r}_{u}+\lambda^{c}_{v})}.
\end{align*}
This can be conveniently expressed as:
\begin{displaymath}
P(E)=\prod_{u,v} \frac{\exp((\lambda^{r}_{u}+\lambda^{c}_{v})\cdot h_{u,v})}{1+\exp(\lambda^{r}_{u}+\lambda^{c}_{v})}.
\end{displaymath}
The parameters $\lambda^{r}_{u}$ and $\lambda^{c}_{v}$ can be computed efficiently.
For a prior belief on the overall density, every edge probability $p_{u,v}$ simply equals the assumed density.

\textit{Type (iii): Additional prior beliefs on densities between bins.}
We can partition nodes in an attributed graph into bins according to their value for a particular attribute. For example, nodes representing people in a university social network can be partitioned by class year. Then expressing prior beliefs regarding the edge density between two bins is possible. This would allow the data analyst to express, for example, an expectation about the probability that people in class year $y_1$ is connected to those in class year $y_2$. If the analyst believes that people in different class years are less likely to connect with each other, the discovered pattern would end up being more informative and useful as it contrasts more with this kind of belief. As shown by Adriaens et al. \cite{florian}, the resulting background distribution is also a product of Bernoulli distributions, one for each of the random variable $h_{u,v}\in \{0,1\})$:
\begin{displaymath}
P(E)=\prod_{u,v}\frac{\exp((\lambda^{r}_{u}+\lambda^{c}_{v}+\gamma_{k_{u,v}})\cdot h_{u,v})}{1+\exp(\lambda^{r}_{u}+\lambda^{c}_{v}+\gamma_{k_{u,v}})},
\end{displaymath}
where $k_{u,v}$ indexes the block formed by the intersecting part of two bins which vertex $u$ and $v$ belong to correspondingly, $\lambda^{r}_{u}$ ,$\lambda^{c}_{v}$ and $\gamma_{k_{u,v}}$ are efficiently computable parameters.
Note that the background distribution can model a prior belief simultaneously for the edge densities between bins resulting from multiple partitions.

\subsubsection*{Updating the background distribution\label{subsec: updatebg}}
Upon being represented with a pattern, the background distribution should be updated to reflect the data analyst's newly acquired knowledge. The beliefs attached to any value for the data that does not contain the pattern should become zero. In the present context, once we present a pattern $(W_1,W_2,I,k)$ to the analyst, the updated background distribution $P'$ should be such that  $\phi_W(E)\geq k_W$ (if $I=0$) or  $\phi_W(E)\leq k_W$ (if $I=1$) holds with probability one, where  $\phi_W(E)$ denotes a function counting the number of edges between $\varepsilon(W_1)$ and $\varepsilon(W_2)$.  By De Bie \cite{debie2011a}, it was argued to choose $P'$ as the \emph{I-projection} of the previous background distribution onto the set of distributions consistent with the presented pattern. Then Van Leeuwen et al. \cite{vanLeeuwen2016} showed that the resulting $P'$ is again a product of Bernoulli distribution:
\begin{align*}
P'(E)&=\prod_{u,v} {p'_{u,v}}^{h_{u,v}}\cdot (1-p'_{u,v})^{1-h_{u,v}}\\
\text{where}\quad  p'_{u,v}&= \left \{ \begin{array}{ll}
						p_{u,v}\quad\ \text{if}\quad\neg \big(u \in \varepsilon(W_1), v\in \varepsilon(W_2) \big),\\
						\frac{p_{u,v}\cdot\exp(\lambda_W)}{1-p_{u,v}+p_{u,v}\cdot\exp(\lambda_{W})}\quad \text{otherwise.}
						\end{array}
					\right.
\end{align*}
How to compute $\lambda_{W}$ is also given in \cite{vanLeeuwen2016}.

\subsection{The subjective interestingness measure}\label{sec:si}

\ptitle{The Information Content (IC)}
Given a pattern $(W_1,W_2,I,k_W)$, and a background distribution defined by $P$, the probability of the presence of the pattern is the probability of getting $k_W$ or more (for $I=0$), or fewer than $k_W$ (for $I=1$) successes in $n_W$ trials with possibly different success probabilities $p_{u,v}$. 
While it is impractical to compute these probabilities exactly, using the same approach as Van Leeuwen et al. \cite{vanLeeuwen2016} they can be tightly upper bounded using the general Chernoff/Hoeffding bound \cite{
hoeffding1963}, as follows:
\begin{displaymath}
\bPr{[(W_1,W_2,I=0,k_W)]}\leq\mbox{exp}\bigg(-n_W\bKL\bigg(\frac{k_W}{n_W}\parallel p_W\bigg)\bigg),
\end{displaymath}
\begin{align*}
\bPr{[(W_1,W_2,I=1,k_W)]} &\\
\leq\mbox{exp}\bigg(&-n_W\bKL\bigg(1-\frac{k_W}{n_W}\parallel 1-p_W\bigg)\bigg),
\end{align*}
where
$p_W =\frac{1}{n_W}\sum_{u \in \varepsilon(W_1), v \in \varepsilon(W_2)}p_{u,v}.$ 

$\bKL\bigg(\frac{k_W}{n_W}\parallel p_W\bigg)$ is the Kullback-Leibler divergence between two Bernoulli distribution with success probabilities  $\frac{k_W}{n_W}$ and $p_W$ respectively. Note that:
\begin{align*}
  \bKL\big(\frac{k_W}{n_W}\parallel p_W\big)&=\bKL\big(1-\frac{k_W}{n_W}\parallel 1-p_W\big),\\
  &=\frac{k_W}{n_W}\log \big(\frac{k_W/n_W}{p_W}\big)+\\
  &\quad\big(1-\frac{k_W}{n_W}\big)\log \big(\frac{1-k_W/n_W}{1-p_W} \big).
\end{align*}
We can thus write:
\begin{displaymath}
\bPr{[(W_1,W_2,I,k_W)]}\leq\mbox{exp}\bigg(-n_W\bKL\bigg(\frac{k_W}{n_W}\parallel p_W\bigg)\bigg).
\end{displaymath}
The IC is the negative log probability of the pattern being present under the background distribution:
\begin{align}
    \mbox{IC}[(W_1,W_2,I,k_W)]&=-\log (\bPr{[(W_1,W_2,I,k_W)]}), \nonumber\\
    &\geq n_W\bKL\bigg(\frac{k_W}{n_W}\parallel p_W\bigg).
\end{align}

\ptitle{The Description Length (DL)}
A pattern with larger IC is more informative. Yet, sometimes it is harder for the analyst to assimilate as its description is more complex. A good SI measure should trade off IC with DL.
The DL should capture the length of the description needed to communicate a pattern. Intuitively, the cost for the data analyst to assimilate a description $W$ depends on the number of selectors in $W$, i.e., $|W|$. Let us assume communicating each selector in a description $W$ has a constant cost of $\alpha$ and the cost for $I$ and $k_W$ is fixed as $\beta$\footnote{In all our experiments, we use $\alpha=0.3, \beta=0.5.$}. The total description length of a pattern $(W_1,W_2,I,k_W)$ can be written as:
\begin{equation}
\mbox{DL}[(W_1,W_2,I,k_W)]=\alpha (|W_1| + |W_2|)+\beta.
\end{equation}

\ptitle{The Subjective Interestingness (SI)}
Putting the IC and DL together finally yields the SI:
\begin{align}
\label{eq: SI}
    \mbox{SI}[(W_1,W_2,I,k_W)] &= \frac{ \mbox{IC}[(W_1,W_2,I,k_W)]}{\mbox{DL}[(W_1,W_2,I,k_W)]}, \nonumber \\
&=\frac{n_W\bKL\bigg(\frac{k_W}{n_W}\parallel p_W\bigg)}{\alpha (|W_1| + |W_2|)+\beta}.
\end{align}

\section{Algorithm\label{sec: algorithm}}

This section describes the algorithm for obtaining a set of interesting patterns. Since the proposed SI interestingness measure is more complex than most objective measures, heuristic search strategies are inevitable for tractability, as described next. 

\subsection{Beam search\label{subsec: beam}}

For mining single-subgroup patterns, we applied a classical heuristic search strategy over the space of descriptions---the beam search. The general idea is to only store a certain number (called the \emph{beam width}) of best partial description candidates of a certain length (number of selectors) according to the SI measure, and to expand those next with a new selector. This is then iterated. This approach is standard practice in subgroup discovery (used e.g. in Cortana \cite{cortana} 
and pysubgroup \cite{pysubgroup}).

\subsection{Nested beam search\label{subsec: 2beam}}

To search for the bi-subgroup patterns, however, a traditional beam search over both $W_1$ and $W_2$ simultaneously turned out to be more difficult to apply effectively: beams large enough for good quality results turned out to be too demanding.
Instead, a nested beam search strategy, where one beam search is nested into the other, gives good results. Here, the outer beam search explores promising selectors for the description $W_1$, and the inner beam search expands those for $W_2$. Let us denote the width of the outer and inner beam by $x_1$ and $x_2$ respectively. The total number of interesting patterns identified by our algorithm is $x_1\cdot x_2$. To maintain a sufficient diversity among the discovered patterns, we constrain the outer beam to contain at least $x_1$ different $W_1$ descriptions. 
Due to the space limitation, further details are given in the Appendix. 
\subsection{Implementation \label{subsec: imp}}

The implementation builds on \emph{Pysubgroup} \cite{pysubgroup}, a Python package for subgroup discovery implementation
. We integrated our nested beam search algorithm and SI measure into this original interface. A Python implementation of the algorithms and the experiments is available.\footnote{\url{https://bitbucket.org/ghentdatascience/essd_public}} All experiments were conducted on a PC running Ubuntu with i7-7700K 4.20GHz CPU and 32 GB of RAM.

\section{Experiments\label{sec: experiments}}

We evaluate our methods on three real-world networks. In the following, we first describe the datasets (Sec.~\ref{subsec: data}). Then we discuss the properties of the discovered patterns (single-subgroup patterns in Sec.~\ref{subsec: single_results} and bi-subgroup patterns in Sec. \ref{subsec: bi_results}), with a purpose to evaluate various aspects of our proposed SI measure. In addition, scalability evaluation for both cases is given.

\subsection{Data \label{subsec: data}}

For our experiments we used four datasets. Data size statistics are given in Table~\ref{tab: data}.

\noindent\ptitle{\textit{Caltech36} and \textit{Reed98}} Two Facebook social networks from the Facebook100 \cite{facebook100} data set, gathered in September 2005: one for Caltech Facebook users, and one for Reed University. Node attributes describe the person's status (faculty or student), gender, major, minor, dorm/house, graduation year, and high school.

\noindent\ptitle{\textit{Lastfm}} \cite{RecSys2011} A 
social network generated from friendships between \texttt{Lastfm.com} users. A list of most-listened musical artists and tag assignments for each user is given in [user, tag, artist] tuples. We took the tags that a user ever assigned to any artist and assigned those to the user as binary attributes expressing a user's music interests.

 \noindent\ptitle{\textit{DblpAffs}} A DBLP\footnote{https://aminer.org/citation} citation network based on a random subset of publications from 20 conferences\footnote{IJCAI, AAAI, ICML, NIPS, ICLR, ICDE, VLDB, SIGMOD, ICDT, PODS, SIGIR, WWW, CIKM, ECIR, KDD, ECML-PKDD, WSDM, PAKDD, ICDM, SDM} selected to cover 4 research areas: Machine Learning, Database, Information Retrieval, and Data Mining. Only papers for which the authors' country (or state, in the USA) of affiliation is available are included. 
The resulting $116$ countries/states are included as binary node attributes, set to $1$ iff one of the paper's authors is affiliated to an institute in that country/state.

\begin{table}
  \caption{Dataset statistics summary}
  \label{tab: data}
  \resizebox{0.49\textwidth}{!}{
  \begin{tabular}{cccccc}
    \toprule
    Dataset&Type&$|V|$&$|E|$&\#Attributes&$|S|$ \\
    \midrule
    \textit{Caltech36} &undirected &762 & 16651& 7&602\\
     \textit{Reed98} &undirected &962 & 18812& 7&748\\
     \textit{Lastfm} &undirected &1892 &12717 &11946 & 23892\\
     \textit{DblpAffs} &directed & 6472 & 3066 & 116 & 232\\
      \bottomrule
\end{tabular}}
\end{table}

\subsection{Results on single-subgroup patterns \label{subsec: single_results}}
First, we analyzed single-subgroup patterns on \textit{Lastfm} using beam search with beam width $20$ and search depth $2$.

\subsubsection{Evaluation of the identified subgroups \label{subsubsubsec: single_sgd}}

When using the SI measure to perform the pattern discovery, the prior belief is on the individual vertex degrees. As a result, single-subgroup patterns' density will not be explainable merely from the individual degrees of the constituent vertices. For \textit{Lastfm}, given its sparsity, incorporating this prior leads to a background distribution with a small average connection probability. In this case, our algorithm tends to identify dense clusters (i.e. $I=0$), as these are more informative. There exist numerous measures objectively quantifying the
interestingness of a dense subgraph community. We make a comparison between our SI measure and some of these objective ones, including the edge density, the average degree, Pool's community score~\cite{Pool2014}, the edge surplus~\cite{tsourakakis2013}, the segregation index~\cite{Freeman1978}, the modularity of a single community~\cite{newman2006, nicosia2009}, the inverse average-ODF (out-degree fraction)~\cite{yang2015} and the inverse conductance. For space limitations, tables with the most interesting patterns w.r.t these measures are put in the Appendix. 
The main findings are summarized here.

Each of those objective measures exhibits a particular bias that arguably makes the attained patterns less useful in practice. The edge density is easily maximized to a value of $1$ simply by considering very small subgraph. That's why the patterns identified by using this measure are all those composed of only $2$ vertices with $1$ connecting edge. In contrast, using the average degree tends to find very large communities, because in a large community there are many other vertices for each vertex to be possibly connected to. Although Pool argued that their measure may be larger for larger communities than for smaller ones, in their own experiments on \textit{Lastfm} as well as in our own results, it yields relatively small communities. As they explained, the reason was \textit{Lastfm}'s attribute data is extremely sparse with a density of merely $0.15\%$. Note the attained patterns from using edge surplus are the same as those using the Pool's measure. Although these two measures are defined in different ways, Pool's measure can be further simplified to a form essentially the same as the edge surplus (shown in the Appendix). Pursuing a larger segregation index essentially targets communities which have much less cross-community links than expected. This measure emphasizes more strongly the number of cross-community links, and yields extremely small or large communities with few inter-edges on \textit{Lastfm}. Using the modularity of a single community tends to find rather large communities representing audiences of mainstream music. The results for the inverse average-ODF and the inverse conductance are not displayed in the Appendix, because the largest values for these two measures can be easily achieved by a community with no edges leaving this community, for which a trivial example is the whole network.

We argue that the attained patterns by applying our SI measure are most insightful, striking the right balance between coverage (sufficiently large) and specificity (not conveying too generic or trivial information). The top one characterises a group of 78 idm (i.e., intelligent dance music) fans. Audiences in this group are connected more frequently than expected, and they altogether only have $496$ connections to those people not into idm, a small number compared to the number of people outside the group (i.e., $1892-78=1814$).

\begin{myrem}
This sort of qualitative comparison was also made on \textit{DblpAffs} (See results in the Appendix), for which the same conclusion as above can be reached. 
\end{myrem}
\subsubsection{Scalability}

\begin{figure}[t]
  \centering
 \includegraphics[width=\linewidth,height=4.1cm]{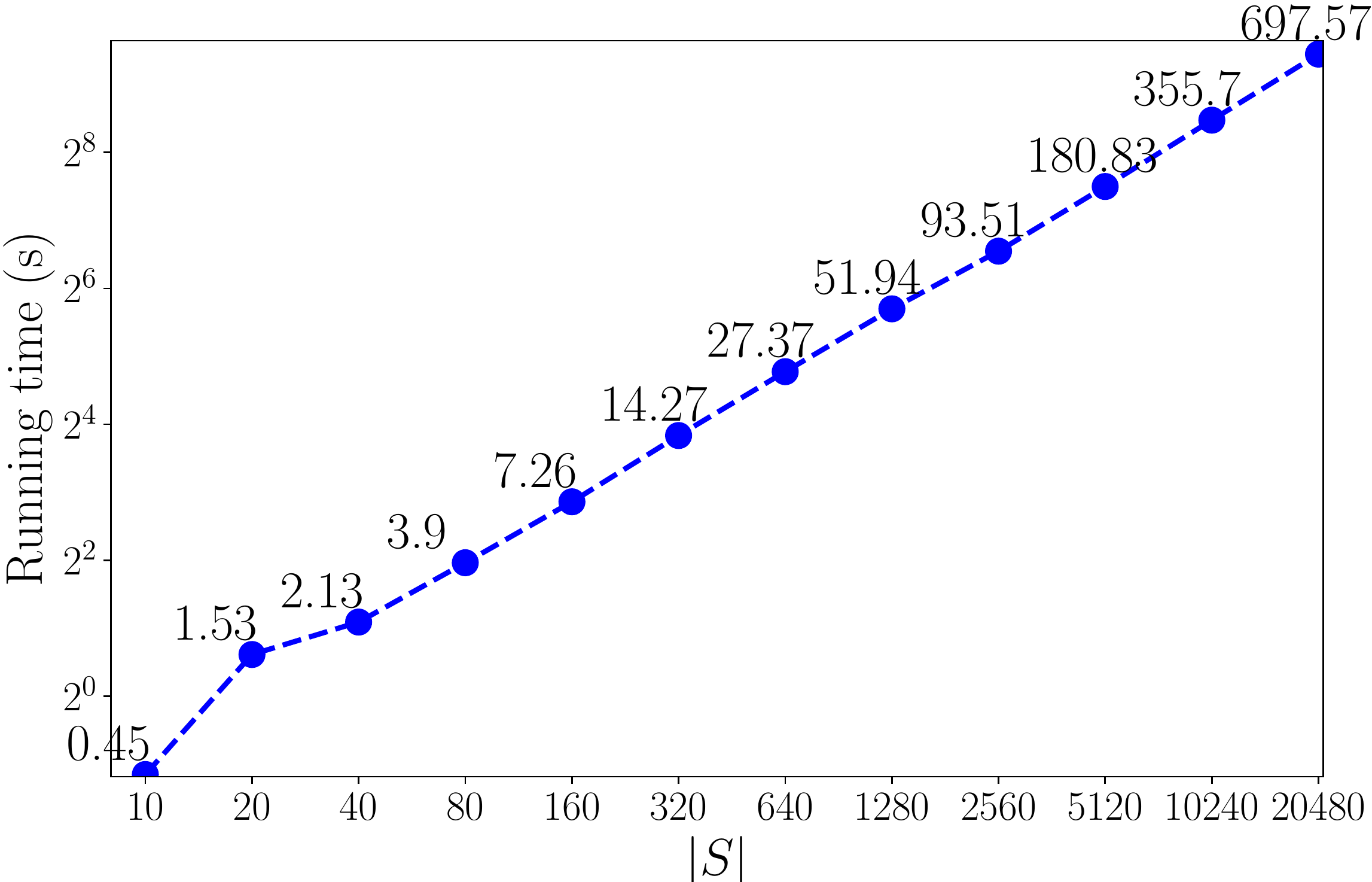}
  \caption{Run time on \textit{Lastfm} for various $|S|$ }
   \label{fig: scalability}
\end{figure}

Fig.~\ref{fig: scalability} illustrates how the algorithm scales w.r.t the number of selectors in the search space (i.e., $|S|$). Both axes are assigned with logarithmic scales with base $2$. It is clear that the run time experiences a linear growth as we double the $|S|$ except a tiny disagreement from the second implementation.

\begin{table*}[tp]
\caption{Varying prior beliefs in \textit{Caltech36} network}
\label{tab: caltech_bi}
\resizebox{\textwidth}{!}{
\begin{tabular}{ccccccccc}
 \toprule
 &Rank &  $W_1$  & $W_2$ & $|\varepsilon(W_1)|$ & $|\varepsilon(W_2)|$&$I$   & $k_W$  & $p_W\cdot n_W$   \\ \toprule\midrule
\multirow{4}{*}{ \textbf{Prior 1}} &1&  year $=2006$  &  year $=2008$  &   153 & 173 & 1 &1346 &2379.10 \\
&2&  status $=$ student $\land $ year $=2008$& status $=$ alumni & 167   & 159  &   1 &842&1783.26\\
 &3&   status $=$ student $\land $ year $=2008$& year $=2006$  & 167   & 153  &   1 &1330&2367.96\\
 &4&  status $=$ student $\land $ year $=2006$& year $=2008$  & 152   & 173  &   1 &1346&2377.53\\ \toprule
\multirow{4}{*}{ \begin{tabular}[c]{@{}l@{}}\textbf{Prior 1}\\+ \textbf{Prior 2}\end{tabular}} &1&  dorm/house $=169$ & dorm/house $=171$  &   99 &  67  & 1 &194 &569.56  \\
 &2&  dorm/house $=169$ & dorm/house $=166$  & 99   & 70  &   1 &237&620.42\\
&3& dorm/house $=169$  &  dorm/house $=172$& 99   & 91  & 1 & 319  &706.65 \\
 &4& dorm/house $=169$  &  dorm/house $=170$& 99   & 87  & 1 & 300 &646.04  \\
  \toprule
 \multirow{4}{*}{\begin{tabular}[c]{@{}l@{}l@{}} \textbf{Prior 1}\\+ \textbf{Prior 2}\\+ \textbf{Prior 3}\end{tabular}}&1&status $=$ student $\land $ year $=2004$& year $=2008$ & 3   & 173  &   0 &108&25.23\\
 &2& status $=$ student $\land $ year $=2004$&  year $=2008$ $\land $ minor $=0$  & 3   & 114  &   0 &71&15.67\\
 &3& status $=$ student $ \land $ year $=2004$&  year $=2008$ $\land $ gender $=$ male& 3   & 116  & 0 & 71 &16.97  \\
 &4&status $=$ student $\land$ dorm/house $=166 $& status $=$ alumni $\land$ high school $=19445 $& 53   & 1& 0 & 51 &17.52  \\
  \bottomrule
\end{tabular}}
\end{table*}

\subsection{Results on bi-subgroup patterns \label{subsec: bi_results}}

To identify bi-subgroup patterns, we applied the nested beam search with $x_1 = 8, x_2=6$, and $D=2$. Moreover, we constrain the target descriptions $W_1$ and $W_2$ to include at least one common attribute but with various values, so that the corresponding pair of subgroups $\varepsilon(W_1)$ and $\varepsilon(W_2)$ do not overlap with each other. Under this setting, the attained patterns are more explainable, and the results are easier to evaluate.

\subsubsection{Evaluation of the SI measure \label{subsubsec: eval_SI}}

The evaluation of the SI measure addresses two questions:
\begin{itemize}
\item Is the SI truly subjective, in the sense of being able to consider data analyst's prior beliefs? (Task 1)
\item How can optimizing SI help avoid redundancy in the resulting patterns from an iterative mining? (Task 2)
\end{itemize}

\noindent\ptitle{Task 1: The effects of different prior beliefs, and a subjective evaluation}
We consider different prior beliefs, in search for bi-subgroup patterns w.r.t our SI measure on \textit{Caltech36} and \textit{Reed98}. The 4 most subjectively interesting patterns under each prior belief are presented in Table~\ref{tab: caltech_bi} (for \textit{Caltech36}) and Table~\ref{tab: reed_bi} (for \textit{Reed98})).
For each pattern, the expected number of edges between $\varepsilon(W_1)$ and $\varepsilon(W_2)$ w.r.t the background distribution (i.e., $p_W\cdot n_W$) is also displayed.

\ptitle{Prior beliefs on the individual vertex degrees} We first incorporated prior belief on the individual vertex degree (i.e. Prior 1). 
In general, the identified patterns belong to knowledge commonly held by people, and are not useful.
The top 4 patterns on \textit{Caltech36} all reveal people graduating in different years rarely know each other (rows for Prior 1 in Table~\ref{tab: caltech_bi}), in particular between ones in class of 2006 and ones in class of 2008 (indicated by the most interesting pattern).
Although $W_2$ of the second pattern (i.e., \emph{status $=$ alumni}) does not contain the attribute graduation year, it implicitly represents people who had graduated in former year. 
For \textit{Reed98}, the discovered patterns under Prior 1 also express the negative influence of different graduation years on connections (rows for Prior 1 in Table~\ref{tab: reed_bi}). 

\ptitle{Prior beliefs on particular attribute knowledge} We then incorporated prior beliefs on the densities between bins for different graduation years (i.e., Prior 2). All the attained top 4 patterns on \textit{Caltech 36} indicate rare connections between people living in different dormitories, and this is also not surprising.
By additionally incorporating prior beliefs on the dependency of the connectivity probability on the difference in dormitories (i.e., Prior 3), patterns characterizing some interesting dense connections are discovered. 
For instance, the top pattern indicates three people in class of 2004 connect with many in class of 2008. In fact, these three people's graduation had been postponed, as their status is `student' rather than `alumni' in year 2005. Furthermore, the starting year for those 2008 cohort is exactly when these three people should have graduated. Therefore, these two groups had opportunities to become friends. 
The forth pattern indicates an alumni who had studied in a high school knew almost all the students living in a certain dormitory. The reason behind this pattern might be worth investigating, which could be for instance, this alumni worked in this dormitory. 
For \textit{Reed98}, incorporating Prior 1 and Prior 2 provides interesting patterns. 
The top one indicates people living in dormitory 88 are friends with many in dormitory 89. In contrast, what people commonly believe is that people living in different dormitories are less likely to know each other. For an analyst who has such preconceived notion, this pattern is interesting.
Both the fourth and the seventh patterns reveal a certain person knew more people in class of 2009 than expected.

\ptitle{Summary} As the results show, incorporating different prior beliefs leads to discovering different patterns that strongly contrast with these beliefs. Our SI measure can quantify the interestingness subjectively.


\begin{table*}[t]
\caption{Varying prior beliefs in \textit{Reed98} network}
\label{tab: reed_bi}
\resizebox{\textwidth}{!}{
\begin{tabular}{M{14mm}M{5mm}M{58mm}M{30mm}M{8mm}M{8mm}M{4mm}M{4mm}M{15mm}}
 \toprule
 &Rank &  $W_1$  & $W_2$ & $|\varepsilon(W_1)|$ & $|\varepsilon(W_2)|$&$I$   & $k_W$  & $p_W\cdot n_W$   \\ \toprule\midrule
 \multirow{4}{*}{ \textbf{Prior 1}}&1&  year $=2008$  &  year $=2005 $  &   209 & 117 & 1 &495 &1401.97 \\
&2& year $=2007$  &  year $=2009 $  &  165 & 158& 1 &112 &661.41 \\
 &3& status $=$ student $\land $ year $=2008$ &year $=2005$  & 209   & 117  &   1 &495&1401.97\\
& 4& year $=2008$  &  year $=2006 $  &   209 & 131 & 1 &765 & 1643.38\\ \toprule
 \multirow{5}{*}{\begin{tabular}[c]{@{}l@{}} \textbf{Prior 1}\\+\textbf{Prior 2} \end{tabular}}&1&   dorm/house $=89$ & dorm/house $=88$  &   23 &  37  & 0 &188 &68.80  \\
&2&  dorm/house $=89$  $\land$ status $=$ student & dorm/house $=88$  &   22 &  37  & 0 &188 &68.45  \\
&3&   dorm/house $=88$  $\land$ status $=$ student & dorm/house $=89$  &   36&  23  & 0 &183 &65.47 \\
 &4&  dorm/house $=111$  $\land$ year $=0 $&  year $= 2009$ & 1   & 158& 0 & 24 &0.66  \\
  &7&  dorm/house $=96$  $\land$ year $=2005 $ &  year $= 2009$ & 1   & 158& 0 & 12 &0.07  \\
  \bottomrule
\end{tabular}}
\end{table*}

\noindent\ptitle{Task 2: Evaluation on the iterative pattern mining}
Our method is naturally suited for iterative pattern mining, in a way to incorporate the newly obtained pattern into the background distribution for subsequent iterations. For this task, we used \textit{DblpAffs} and \textit{Lastfm} dataset. Results for \textit{Lastfm} are displayed and discussed in the Appendix. Here we only analyze the results on \textit{DblpAffs}. Table~\ref{tab: dblpAffs_bi} displays top 3 patterns found in each of the four iterations on \textit{DblpAffs}.

\ptitle{Iteration 1} Initially, we incorporated prior on the overall graph density. The resulting top pattern indicates papers from institutes in USA seldom cite those from other countries. 

\ptitle{Iteration 2} After incorporating the top pattern in iteration 1, a set of dense patterns were identified. All the top 3 patterns reveal a highly-cited subgroup of papers whose authors are affiliated to institutes in California and New Jersey.
This is possible as many of the world's largest high-tech corporations and reputable universities are located in this region. 
Examples include Silicon valley, Stanford university in CA, NEC Laboratories, AT\&T Laboratories in NJ, among others.  

\ptitle{Iteration 3} The top 3 patterns in iteration 3 reveal that papers from authors with Chinese affiliations are rarely cited by papers with authors from other countries. However, they are frequently cited by papers with Chinese authors, as indicated by our identified top single-subgroup pattern in \textit{DblpAffs} (see supplement). This indicates researchers with Chinese affiliations are surprisingly isolated, the reason of which might be interesting to investigate.
 
\ptitle{Iteration 4} The top patterns in iteration 4 reveal that papers from institutions in Washington state are highly cited by others, in particular by papers from California. Closer inspection revealed that the majority of these papers are written by authors from Microsoft Corporation and the University of Washington. 

\ptitle{Summary}  By incorporating the newly attained patterns into the background distribution for subsequent iterations, our method can identify patterns which strongly contrast to this knowledge. This results in a set of patterns that are not redundant and highly surprising to the data analyst. Note this does not means we restrict patterns in different iterations not to be associated with each other. In fact, overlapping could happen when this is informative.

\begin{table*}
\caption{Top 3 discovered bi-subgroup patterns of each iteration in \textit{DblpAffs} network}
\label{tab: dblpAffs_bi}
 \resizebox{\textwidth}{!}{
\begin{tabular}{M{20mm}M{5mm}M{45mm}M{48mm}M{8mm}M{8mm}M{3mm}M{4mm}M{15mm}}
 \toprule
 & Rank &  $W_1$  & $W_2$ & $|\varepsilon(W_1)|$ & $|\varepsilon(W_2)|$&$I$   & $k_W$  & $p_W\cdot n_W$   \\ \toprule\midrule
\multirow{3}{*}{ \textbf{Iteration 1}}  &1&  USA $=1$  &  USA $=0$  &  3132 &  3340 & 1 &335 &765.827  \\
& 2& USA$=1$ $\land $ China $=0$  & USA $=0$  & 2969    & 3340 & 1 & 288&725.970\\
& 3& USA$=1$ $\land $ Australia $=0$  & USA $=0$  & 3092   & 3340 & 1 & 320&756.046\\
 \toprule
\multirow{3}{*}{ \textbf{Iteration 2}} & 1&  NJ (New Jersey) $=0$ &  NJ $=1$ $ \land $ CA (California) $=1$  &   6262 &  15  & 0 &93 &6.909  \\
&2&  CA $=0$ &  NJ$=1$ $ \land $ CA $=1$  &   5584 &  15  & 0 &86 &6.132  \\
&3&  NJ$=1$ $ \land $ Israel $=0$  &   NJ$=1$ $ \land $ CA $=1$& 6153   & 15  & 0 & 93 &6.757  \\
  \toprule
 \multirow{3}{*}{ \textbf{Iteration 3}} &1&  China $=0$ &  China $=1$ &   5599 &  873  & 1 &144 &271.022  \\
 &2& China $=0$ &  China $=1$ $\land$ IL (Illinois) $=0$&   5599 &  861  & 1 &128 &266.103  \\
&3& China $=0$ $\land$ USA $=0$ &  China $=1$ &   2630 &  873  & 1 &64 &168.086  \\
   \toprule
\multirow{3}{*}{ \textbf{Iteration 4}}  &1&  CA $=1$ &  CA $=0$ $\land$ WA $=1$&   888 &  184  & 0 &55 &11.726  \\
& 2&  WA $=0$ &WA $=1$ & 6254  & 218  &  0 &182&97.776\\
 &3& CA $=1$ $\land$ TX (Texas) $=0$  &  CA $=0$ $\land$ WA $=1$& 876   & 184  & 0 & 55 &11.568  \\
 \bottomrule
  \end{tabular}}
\end{table*}

\subsubsection{Evaluation on the run time}

The run time of the nested beam search on each dataset, as well as the $|S|$ and $|V|$ statistics are listed in Table~\ref{tab: runtime_bi}. The influence of the $|S|$ and $|V|$ on the run time is evident.
\begin{table}
  \caption{Run time of bi-subgroup pattern mining}
  \label{tab: runtime_bi}
   \resizebox{0.35\textwidth}{!}{
\begin{tabular}{cccc}
    \toprule
    Dataset&$|S|$&$|V|$&Run time (s) \\
    \midrule
    \textit{Caltech36} &602 &762 &6855.52\\
    \textit{Reed98} &748 & 962 &10692.83\\
    \textit{Lastfm} &200 &1892 &5954.50\\
    \textit{DblpAffs} &232 &6472 &10015.70\\
         \bottomrule
\end{tabular}}
\end{table}

\section{Related work\label{sec: relatedwork}}

Real-life graphs often have attributes on the vertices. Pattern mining considering both structural aspect and attribute information promises more meaningful results, and thus has received increasing research attention.
The problem of mining cohesive patterns was introduced by Moser et al. \cite{Moser2009}. They define a cohesive pattern as a connected subgraph whose edge density exceeds a given threshold, and vertices exhibit sufficient homogeneity in the attribute space. Gunnemann et al. \cite{Gunnemann2010} propose to combine subspace clustering and dense subgraph mining. The former technique is to determine set of nodes that are highly similar according to their attribute values, and the latter is to pursue the cohesiveness of the attained subgraph. Mougel et al. \cite{Mougel2010} compute all maximal homogeneous clique sets that satisfy some user-defined constraints. All these work emphasizes on the graph structure and consider attributes as complementary information.

Rather than assuming attributes to be complementary, descriptive community mining, introduced by Pool et al. \cite{Pool2014}  aims to identify cohesive communities that have a concise description in the vertices' attribute space. They propose cohesiveness measure, which is based on counting erroneous links (i.e., connections that are either missing or obsolete w.r.t the `ideal' community given the induced subgraph). To a limited extent, their method can be driven by user's domain-specific background knowledge, and specifically, it is a preliminary description or a set of nodes that are expected to be part of a community. Then the search is triggered by those seed candidates. Our proposed SI, in contrast, is more versatile in a sense that allows incorporating more general background knowledge. Galbrun et al. \cite{Galbrun2014} proposes a similar target to Pool et al.'s, but relies on a different density measure, which is essentially the average degree. 
Atzmueller et al. \cite{Atzmueller2016} introduce description-oriented community detection. They apply a subgroup discovery approach to mine patterns in the description space so it comes naturally that the identified communities have a succinct description.


All previous works quantify the interestingness in an objective manner, in the sense that they can not consider a data analyst's prior beliefs and thus operate regardless of context. 
Also, all previous works focus on a set of communities or dense subgraphs, overlooking other meaningful structures such as a sparse or dense subgraph between two different subgroups of nodes.

\section{Conclusion\label{sec: conclusion}}

We presented a method to identify patterns in the form of (pairs of) subgroups of nodes in a graph,
such that the density of (the graph between) those node subgroups is interesting.
Here, `interesting' is quantified in a subjective manner, with respect to a flexible type of prior knowledge about the graph the analyst may have, including insights gained from previous patterns.

Our approach improves upon the interestingness measures used in prior work on subgroup discovery for dense subgraph mining in attributed graphs,
and generalizes it in two ways: in identifying not only dense but also sparse subgraphs,
and in describing the density between subgroups that may differ from each other.

The empirical results show that the method succeeds in taking into account prior knowledge in a meaningful way,
and is able to identify patterns that provide genuine insight into the high-level network's structure.

\section{Appendix}
This section consists of:
\begin{itemize}
\item[--] The pseudo-code and the notation description of the nested beam search in Sec.~\ref{subsec: 2beam} of the main paper;
\item[--] Top 4 single-subgroup patterns w.r.t the SI on \textit{Lastfm} and \textit{DblpAffs} for Sec.~\ref{subsubsubsec: single_sgd} of the main paper;
\item[--] The description of those baseline objective measures we used for a comparative evaluation for Sec.~\ref{subsubsubsec: single_sgd} of the main paper;
\item[--] Top 4 single-subgroup patterns w.r.t other objective measures on \textit{Lastfm} and \textit{DblpAffs} for Sec.~\ref{subsubsubsec: single_sgd} of the main paper;
\item[--] Evaluation on the iterative pattern mining on \textit{Lastfm} Dataset for the task 2 in Sec.~\ref{subsubsec: eval_SI} of the main paper.
\end{itemize}

\subsection{The pseudo-code and the notation description of the nested beam search}
The detailed procedure for the nested beam search is shown in  Algorithm~\ref{algo: 2beam}, as well as its related notations displayed in Table~\ref{tab: no}.

\begin{table}[h]
  \caption{Notations for Algorithm~\ref{algo: 2beam}}
  \label{tab: no}
  \resizebox{0.5\textwidth}{!}{
  \begin{tabular}{P{0.08\textwidth}P{0.36\textwidth}}
    \toprule
    Notation&Description\\
    \midrule
    Beam & The outer beam storing best description pairs $(W_1, W_2)$ during the search.\\
    InnerBeam & The inner beam only storing best descriptions $W_2$.\\
    $x_1$ & The minimum number of different descriptions $W_1$ contained in the beam.\\
    $x_2$ & The inner beam width.\\
   $D$ & The search depth (i.e., maximum number of selectors combined in a description).\\
  \bottomrule
\end{tabular}}
\end{table}

\IncMargin{1em}
\begin{algorithm2e*}
\caption{Subjectively Interesting BiSubgroup Pattern Mining}\label{algo: 2beam}

\SetKwFunction{AddIfRequired}{AddIfRequired}
\SetKwInOut{Input}{input}\SetKwInOut{Output}{output}

\Input{Graph $G=\{V,E, A\}$, $x_1$,  $x_2$, $D$}
\Output{Beam}
\BlankLine

$S \leftarrow$ the set of all selectors to build descriptions from\;
Beam $\leftarrow \{\emptyset\}$ \;
$d_1 \leftarrow 0$,  $d_2 \leftarrow 0$\;
\While(\tcp*[h]{The outer search}){$d_1< D$}{
$\mathbb{C}_1 \leftarrow$ all the $W_1$ candidate in Beam\;
\For(\tcp*[h]{Expand on $W_1$ candidates}){ $C_1 \in \mathbb{C}_1$}{
\For{$s_1 \in S$}{
$Z_1 \leftarrow C_1\land s_1$\;
InnerBeam $\leftarrow\{ \emptyset\}$\;
\While(\tcp*[h]{The inner search}){$d_2< D$}{
$\mathbb{C}_2 \leftarrow$ all the $W_2$ candidates in InnerBeam\;
\For(\tcp*[h]{Expand $W_2$ candi.}){ $C_2 \in \mathbb{C}_2$}{
\For{$s_2 \in S$}{

$Z_2 \leftarrow C_2\land s_2$\;
$k_W \leftarrow $ the number of edges between nodes $\varepsilon(Z_1)$ and $\varepsilon(Z_2)$\;

\tcp{compute SI of the pattern $(Z_1, Z_2, I, k_W)$ using Eq.~3.3}
 $\text{si}' \leftarrow \text{SI}[(Z_1, Z_2, I, k_W)]$\;
 \tcp{Add (si$',Z_2)$ to the InnerBeam if InnerBeam contains less than $x_2$ elements or replace the tuple with the smallest SI in InnerBeam if $\text{si}'$ is larger than that value}
 InnerBeam $\leftarrow$ \AddIfRequired{\emph{InnerBeam}, (\emph{si}$',Z_2)$, $x_2$}\;

}
}
$d_2 \leftarrow d_2+1$\;
}
\For{(\emph{si},$Z)\in$ \emph{InnerBeam}}{
\tcp{Add (si$,Z_1, Z)$ to the Beam if the number of various $W_1$ descriptions in Beam is less than $x_1$ or replace the tuple with the smallest SI if $\text{si}$ is larger than that value}
Beam $\leftarrow$ \AddIfRequired{\emph{Beam}, (\emph{si}$, Z_1,Z)$, $x_1$}\;
}
}
}
$d_1 \leftarrow d_1+1$\;
}
\end{algorithm2e*}
\DecMargin{1em}

\subsection{Top 4 single-subgroup patterns w.r.t the SI on \textit{Lastfm} and \textit{DblpAffs}}

In the following, 4 most subjectively interesting single-subgroup patterns are presented. (Table~\ref{tab: top4_SI_lastfm} for \textit{Lastfm}, and Table~\ref{tab: top4_SI_dblpAffs} for \textit{DblpAffs}). For each pattern $(W, I, k_W)$, we also display other related statistics including its SI value, $p_W\cdot n_W$ and \#inter-edges. $p_W \cdot n_W$ is the expected number of edges within $\varepsilon(W)$ w.r.t the background distribution and  \#inter-edges is the number of edges between $\varepsilon(W)$ and $V \setminus \varepsilon(W)$.

\begin{table*}[!]
\caption{Top 4 discovered single-subgroup patterns in \textit{Lastfm} network using our SI measure}
\label{tab: top4_SI_lastfm}
 \resizebox{0.6\textwidth}{!}{
\begin{tabular}{ccccccc}
 \toprule\midrule
 SI &  $W$  & $|\varepsilon(W)|$  &$I$ & $k_W$  & $p_W\cdot n_W$ & \#inter-edges   \\ \toprule
355.533&  idm $=1$  &  78  &   0 &  96 & 8.929 &496   \\
316.725&  heavy metal $=1$  &  165   &   0 &  220 & 60.040 &1322   \\
296.894& synthpop $=1$  &  131   &   0 &  208 & 57.320 &1307   \\
286.061&  new wave $=1$  &  191   &   0 &   292 & 104.005 &1731   \\
  \bottomrule
    \end{tabular}
}
\end{table*}
\begin{table*}[!]
\caption{Top 4 discovered single-subgroup patterns in \textit{DblpAffs} network using our SI measure}
\label{tab: top4_SI_dblpAffs}
  \resizebox{0.7\textwidth}{!}{
\begin{tabular}{ccccccc}
 \toprule\midrule
 SI &  $W$  & $|\varepsilon(W)|$  &$I$ & $k_W$  & $p_W\cdot n_W$ & \#inter-edges   \\ \toprule
88.211&   China $ =1$  &  873  &   0 & 179 & 63.197 &566   \\
65.187&  China $=1$ $\land$ IN (Indiana)  $=0$  & 869   &   0 &  179 & 62.581 &561   \\
65.037&  China $=1$ $\land$ Italy  $=0$  & 870   &   0 &  179 & 62.670 &561   \\
65.010&  China $=1$ $\land$ Denmark  $=0$  & 870   &   0 &  179 & 62.686 &562   \\
  \bottomrule
    \end{tabular}
  }
\end{table*}
\subsection{Baseline objective interestingness measures for comparison  }
In the following, those objective interestingness measures we used for a comparative evaluation are described in Table~\ref{tab: othermeasures}. For a given attributed graph $G=\{V,E,\Lambda\}$, and a community induced by a description $W$ such that $\varepsilon(W)\in V$, we use these additional notations in the column of mathematical definition in Table~\ref{tab: othermeasures}. $d(u)$ denotes the degree of node $u \in V$. $\overline{d}_W(u)$ denotes the inter-degree of node $u \in \varepsilon(W)$, specfically, $\overline{d}_W(u) := |\{(u,v)\in E: v \in V \setminus \varepsilon(W)\}| $. 

We consider undirected graphs for the sake of presentation and consistency with most literature. However, we note that the generalization to directed graphs is straightforward. 

\subsection{Top 4 single-subgroup patterns w.r.t other objective measures on \textit{Lastfm} and \textit{DblpAffs}}
In the following, top 4 single-subgroup patterns w.r.t other objective measures are displayed. (Table~\ref{tab: top4_others_lastfm} for \textit{Lastfm}, and Table~\ref{tab: top4_others_dblpAffs} for \textit{DblpAffs}).

\subsection{Evaluation on the iterative pattern mining on \textit{Lastfm}}
Table~\ref{tab: lastfm_bi} displays the top 3 patterns found in each of the five iterations on the \textit{Lastfm}. The description search space is built based on only 100 most frequently used tags, that means,  $|S|=100\times2$.

\ptitle{Iteration 1} Initially, we incorporate prior belief on individual vertex degree. The attained most interesting pattern reflects a conflict between aggressive heavy metal fans and mainstream pop lovers who do not listen to heavy metal at all.

\ptitle{Iteration 2} After incorporating the top pattern identified in iteration 1, what comes top is the one expressing again a conflict between mainstream and non-mainstream music preference, but another kind (i.e., pop with no indie, and experimental with no pop). Also, we can notice only the second pattern for the iteration 1 is remained in the iteration 2 top list but with a lower rank as third. The interestingness of any sparse pattern associated with the newly incorporated one under the updated background distribution is expected to decrease, as the data analyst's would not feel surprised about such pattern.

\ptitle{Iteration 3} In iteration 3, our method tends to identify some interesting dense patterns, mainly related to synth pop and new wave genres. The top one states synth pop fans frequently connect with many people listening to new wave but not synth pop. This pattern appears fallacious at the first glance. Nevertheless, synth pop is a subgenre of new wave music. Also, the latter group may listen to synth pop but they use a different tag `synthpop' instead of `synth pop', as there are even $102$ audience only tag synth pop as 'synthpop' (see the third patten). Hence, this pattern makes sense as it describes dense connections between two groups which resemble each other. 

\ptitle{Iteration 4} The top 3 patterns in iteration 4 all express negative associations between new wave and some sort of catchy mainstream music (eg. pop,  rnb, or hip-hop, among several others).

\ptitle{Iteration 5} Once we incorporate the most interesting one, patterns characterizing some positively associated genres stand out. For example, the top one in iteration 5 indicates instrumental audience are friends with many ambient audience who doesn't listen to instrumental music. These two genres are not opposite concepts and share many in common (e.g., recordings for both do not include lyrics). Actually, ambient music can be regarded as a slow form of instrumental music.

\ptitle{Summary} By incorporating the newly attained patterns into the background distribution for subsequent iterations, our method can identify patterns which strongly contrast to this knowledge. This results in  a set of patterns that are not redundant and are highly surprising to the data analyst. Note this does not means we restrict patterns in different iterations not to be associated with each other. In fact, overlapping could happen when this is informative.

\clearpage

\begin{table*}[h]
  \caption{Existing measures for a comparison}
  \label{tab: othermeasures}
  \resizebox{\textwidth}{!}{
  \begin{tabular}{ccc}
    \toprule\midrule
    \textbf{Measure}&\textbf{Description}&\textbf{Mathematical definition} \\
    \midrule
    Edge density & \begin{tabular}[c]{@{}l@{}}the ratio of the number of edges to the \\ number of possible edges in the cluster\end{tabular} & $\frac{2k_W}{|\varepsilon(W)|\cdot(|\varepsilon(W)|-1)}$\\ \midrule
    Average degree &  \begin{tabular}[c]{@{}l@{}}the ratio of the degree sum for all nodes \\ to the number of nodes in the cluster \end{tabular}&$\frac{2k_W}{|\varepsilon(W)|}$ \\ \midrule
   Pool's measure~\cite{Pool2014} & \begin{tabular}[c]{@{}l@{}l@{}l@{}}the reduction in the number of erroneous \\links  between treating each vertex as a \\single community and treating all the vertices \\as a whole \end{tabular}& \begin{tabular}[c]{@{}l@{}}  $\sum_{u\in \varepsilon(W)} d(u) - \bigg(\frac{|\varepsilon(W)|\cdot(|\varepsilon(W)|-1)}{2}-k_W\bigg)$\\$-\#\mbox{inter-edges}=-\frac{|\varepsilon(W)|\cdot(|\varepsilon(W)|-1)}{2}+3k_W$ \end{tabular}\\ \midrule
   Edge Surplus~\cite{tsourakakis2013} &  \begin{tabular}[c]{@{}l@{}l@{}}the number of edges exceeding the expected \\number of edges within the cluster assuming \\each edge is present with the same probability $\alpha$\end{tabular}& $k_W - \alpha \cdot |\varepsilon(W)|\cdot(|\varepsilon(W)|-1) $\\ \midrule
    Segregation index~\cite{Freeman1978} &  \begin{tabular}[c]{@{}l@{}l@{}}the difference between the number of expected \\inter-edges to the number of the observed \\inter-edges, normalized by the expectation \end{tabular}& $1-\frac{\mbox{ \#inter-edges}\cdot|V|(|V|-1)}{2|E||\varepsilon(W)|\cdot(|V|-|\varepsilon(W)|)}$\\ \midrule
   \begin{tabular}[c]{@{}l@{}} Modularity of a single \\community~\cite{newman2006, nicosia2009} \end{tabular}&  \begin{tabular}[c]{@{}l@{}l@{}l@{}}the measure quantifying the modularity \\contribution of a single community based on \\transforming the definition of modularity \\to a local measure  \end{tabular}& $\frac{1}{2|E|}\sum_{u,v\in \varepsilon(W)} \bigg(a_{u,v}-\frac{d(u)\cdot d(v)}{2|E|}\bigg)$\\ \midrule
    \begin{tabular}[c]{@{}l@{}}Inverse Average-ODF \\(out-degree fraction)~\cite{yang2015} \end{tabular}&\begin{tabular}[c]{@{}l@{}l@{}}the inverse of the Average-ODF which is \\based on averaging the fraction of inter-degree\\ and the degree for each node in the cluster \end{tabular}& $1-\frac{1}{|\varepsilon(W)|}\sum_{u\in \varepsilon(W)}\frac{\overline{d}_W(u)}{d(u)}$\\ \midrule
    Inverse Conductance &\begin{tabular}[c]{@{}l@{}l@{}}the ratio of the number of edges \\inside the cluster to the number of edges\\ leaving the cluster\end{tabular}& $\frac{k_W}{\#\mbox{inter-edges}}$\\

      \bottomrule
\end{tabular}}
\end{table*}

\begin{table*}[tp]
\caption{Top 4 discovered single-subgroup patterns in \textit{Lastfm} network using other measures}
\label{tab: top4_others_lastfm}
 \resizebox{0.87\textwidth}{!}{
\begin{tabular}{ccccc}
 \toprule
 Measure &  $W$  & $|\varepsilon(W)|$   & $k_W$  & \#inter-edges   \\  \toprule\midrule
\multirow{4}{*}{ \textbf{Edge Density}} &  1981 songs $=1$  &  2   &  1 & 21   \\
 &  africa $=1$  &  2  &  1 & 76   \\
 &  40s $=1$  &  2  &  1 & 22   \\
 &  early reggae $=1$  &  2  &  1 & 10   \\
  \midrule
\multirow{4}{*}{ \textbf{Average Degree}}& post rock $=0 \land$ post-rock $=0$ & 1783&12181&498\\
& post-rock $=0 \land$ dark ambient $=0$ & 1770&12092&573\\
 & post-rock $=0 \land$ grindcore $=0$ & 1762&12032&634\\
 & post-rock $=0 \land$ technical death metal $=0$ & 1773&12106&560\\
  \midrule
\multirow{2}{*}{\textbf{Pool's community score}}& bionic $=1 \land$ 30 seconds to mars $=0$ & 6 &8&343\\
& bionic $=1 \land$ taylor swift $=0$ & 6&8&343\\
\multirow{2}{*}{\textbf{ or Edge surplus}}& bionic $=1 \land$ latin $=0$ & 6&8 &343\\
& bionic $=1 \land$ spanish $=0$ & 6 &8&343\\
 \midrule
\multirow{4}{*}{\textbf{Segregation Index}}& gluhie 90e$=0 \land $ lithuanian black metal $=1$ & 3 &3 &1 \\
& goddesses$=0 \land $ pagan black metal $=1$ & 3 &3 &1 \\
& gluhie 90e$=0 \land $ pagan black metal$=1$ & 3 &3 &1 \\
& heartbroke$=0 \land $ lithuanian black metal $=1$ & 3 &3 &1 \\
 \midrule
\multirow{2}{*}{\textbf{Modularity of a}}& pop $=1 \land $ new wave $=0$ & 475 &2689 &4913 \\
& pop $=1 \land $ progressive rock $=0$ & 514 &2943 &5083 \\
\multirow{2}{*}{\textbf{single community}}& pop $=1 \land $ experimental $=0$ & 497 &2844 &5083 \\
& pop $=1 \land $ metal $=0$ & 496 &2761 &5067\\
\bottomrule
\end{tabular}}
\end{table*}

\begin{table*}[tp]
\caption{Top 4 discovered single-subgroup patterns in \textit{DblpAffs} network using other measures}
\label{tab: top4_others_dblpAffs}
 \resizebox{0.87\textwidth}{!}{
\begin{tabular}{ccccc}
 \toprule
 Measure &  $W$  & $|\varepsilon(W)|$   & $k_W$  & \#inter-edges   \\  \toprule\midrule
\multirow{4}{*}{ \textbf{Edge Density}} &DE (Delaware) $=1$ $\land$ MD(Maryland) $=1$ &  2   &  1 & 2   \\
 &  DC (District of Columbia) $=1$ $\land$ TX (Texas) $=1$  &  2  &  1 & 6   \\
 &  Netherlands $=1$ $\land$ MA(Massachusetts) $=1$  &  2  &  1 & 3   \\
 &  Netherlands $=1$ $\land$ WA $=1$  &  2  &  1 & 5   \\
  \midrule
\multirow{4}{*}{ \textbf{Average Degree}}& UK $=0$ $\land$ Japan $=0$ & 6038&2882&161\\
& UK $=0$ $\land$ Ireland $=0$ &6234&2975&79\\
 & Japan $=0$ $\land$ Ireland $=0$ & 6191&2952&106\\
 & Sweden $=0$ $\land$ Ireland $=0$ & 6391&3044&22\\
  \midrule
\multirow{2}{*}{\textbf{Pool's community score}} &DE$=1$ $\land$ MD$=1$ &  2   &  1 & 2   \\
 &  DC $=1$ $\land$ TX $=1$  &  2  &  1 & 6   \\
\multirow{2}{*}{\textbf{ or Edge surplus}}& Netherlands $=1$ $\land$ MA $=1$  &  2  &  1 & 3   \\
&  Netherlands $=1$ $\land$ WA $=1$  &  2  &  1 & 5   \\
 \midrule
\multirow{4}{*}{\textbf{Segregation Index}}&AL (Alabama)$=0 $ & 6470 &3066 &0 \\
&AL $=1 $ & 2 &0 &0 \\
&Bulgaria$=0 $ & 6471 &3066 &0 \\
&AS (American Samoa)$=0 $ & 6471 &3066 &0 \\
 \midrule
\multirow{2}{*}{\textbf{Modularity of a}}&China $=0$ $\land$ United States $=1$ & 2969&1173&1203\\
&NY(New York) $=0$ $\land$ United States $=1$ & 2757&1067&1224\\
\multirow{2}{*}{\textbf{single community}}&Singapore $=0$ $\land$ United States $=1$ & 3088&1247&1194\\
&Germany $=0$ $\land$ United States $=1$ & 3077&1262&1191\\
\bottomrule
\end{tabular}}
\end{table*}

\begin{table*}[tp]
\caption{Top 3 discovered bi-subgroup patterns of each iteration in \textit{Lastfm} network}
\label{tab: lastfm_bi}
 \resizebox{\textwidth}{!}{
\begin{tabular}{M{20mm}M{5mm}M{46mm}M{50mm}M{8mm}M{8mm}M{3mm}M{4mm}M{15mm}}
 \toprule
 & Rank &  $W_1$  & $W_2$ & $|\varepsilon(W_1)|$ & $|\varepsilon(W_2)|$&$I$   & $k_W$  & $p_W\cdot n_W$   \\ \toprule\midrule
\multirow{3}{*}{ \textbf{Iteration 1}}  &1&  heavy mental $=1$  &  heavy mental $=0$ $ \land $ pop $=1$  &   165 &  529 & 1 &349 &769.18  \\
& 2&  pop$=1$ $\land $experimental $=0$  & rnb $=0$ $ \land $experimental$=1$  & 497   & 230  &   1 &360&812.78\\
& 3&  pop$=1$ $\land $experimental $=0$  & experimental $=1$& 497   & 247  & 1 & 495 &943.96  \\
 \toprule
\multirow{3}{*}{ \textbf{Iteration 2}} & 1&  pop $=1$ $\land$ indie $=0$ &  pop $=0$ $ \land $experimental $=1$  &   366 &  159  & 1 &103 &369.44  \\
&2&  pop $=1$ $ \land $ alternative $=0$  & pop $=0$ $\land $ experimental $=1$  & 325   & 159  &   1 &84&334.77\\
  &3&  pop $=1$ $ \land $ experimental $=0$  &  rnb $=0$ $ \land $ experimental $=1$& 497   & 230  & 1 & 360 &750.77  \\
  \toprule
 \multirow{3}{*}{ \textbf{Iteration 3}} &1&  synth pop $=1$ &  synth pop $=0$ $ \land$ new wave $=1$  &   54 &  150  & 0 &163 &43.10  \\
 &2&  synth pop $=1$ $ \land $ british $=1$  & new wave $=1$ $ \land $british $=0$  & 26   & 113  &   0 &116&20.71\\
  & 3&  synth pop $=1$  &  synth pop $=0$ $ \land$ synthpop $=1$& 54   & 102  & 0 & 125 &29.64  \\
   \toprule
\multirow{3}{*}{ \textbf{Iteration 4}}  &1&  new wave $=1$ $ \land $ hip-hop $=0 $ & new wave $=0$ $\land$ pop $=1$ &   160  & 475  & 1 &343 &670.74  \\
& 2&  new wave $=1$ $\land $ rnb $=0$  & new wave $=0$ $\land $ pop $=1$  & 170   & 475  &  1 &379&705.43\\
 &3&  new wave $=1$ $\land $ soul $=0$  &  new wave $=0$ $ \land $ pop $=1$& 150   & 475  & 1 & 323 &624.41  \\
\toprule
\multirow{3}{*}{ \textbf{Iteration 5}} & 1&  instrumental $=1$ &  instrumental $=0$ $ \land$ ambient $=1$  &   195 &  144 & 0 &273 &114.62  \\
&2&  electronic $=1 $  & electronic $=0$ $\land $ ambient $=1$  & 167   & 160 &   0 &268&113.66\\
 &3&  progressive metal $=1$  &  progressive metal $=0$ $ \land$ heavy metal $=1$& 99   & 111  & 0 & 128 &34.81  \\
 \bottomrule
  \end{tabular}}
\end{table*}

\clearpage
\bibliographystyle{plain}
\bibliography{cites}

\end{document}